\documentclass[useAMS,usenatbib]{mn2e}

\usepackage{natbib}
\usepackage{graphicx}
\usepackage{amssymb}
\usepackage{color}
\usepackage{caption}
\usepackage{lipsum,graphicx,multicol}
\usepackage{float}

\usepackage{xcolor}

\usepackage{subcaption}

\title[BH spin-induced anisotropic feedback]{AGN anisotropic radiative feedback set by black hole spin}

\author[ ]
{W. Ishibashi$^{1}$\thanks{E-mail: wako.ishibashi@physik.uzh.ch}
\footnotemark[0]\\
\footnotemark[0]\\
$^{1}$Physik-Institut, Universitat Zurich, Winterthurerstrasse 190, 8057 Zurich, Switzerland 
}

\begin{document}

\pdfminorversion=4

\date{Accepted ? Received ?; in original form ? }

\pagerange{\pageref{firstpage}--\pageref{lastpage}} \pubyear{2012}

\maketitle

\label{firstpage}

\begin{abstract}
We consider the impact of anisotropic radiation on the active galactic nucleus (AGN) radiative dusty feedback. The radiation pattern originating from the accretion disc is determined by the central black hole (BH) spin. Here we analyse how such BH spin-induced angular dependence affects the dynamics and energetics of the radiation pressure-driven outflows, as well as AGN obscuration and BH accretion. In addition, we explore the effect of a spatially varying dust-to-gas ratio on the outflow propagation. We obtain two distinct trends for high-spin and low-spin objects, providing a direct connection between anisotropic feedback and BH spin. In the case of maximum spin, powerful quasi-spherical outflows can propagate on large scales, at all inclination angles with fairly uniform energetics. In contrast, in the case of zero spin, only weaker bipolar outflows can be driven in the polar directions. As a result, high BH spins can efficiently clear out the obscuring gas from most directions, whereas low BH spins can only remove dusty gas from the polar regions, hence also determining the overall AGN obscuration geometry. Due to such anisotropic feedback, high BH spins can prevent accretion of gas from most directions (except in the equatorial plane), while low BH spins allow inflows to proceed from a wider range of directions. This may have important implications for the BH growth in the early Universe. Anisotropic radiative dusty feedback, ruled by the BH spin, may thus play a major role in shaping AGN evolution over cosmic time. 
\end{abstract}



\section{Introduction}

Active galactic nucleus (AGN) feedback is generally invoked in co-evolutionary models in order to reproduce the observed galaxy properties, but the actual mechanism driving such AGN feedback is still a source of much debate. In principle, feedback from the central black hole (BH) can be driven by different physical mechanisms such as jets, winds, and radiation \citep[e.g.][and references therein]{Fabian_2012, King_Pounds_2015}. Two main modes of AGN feedback are currently debated: the radiative (or quasar) mode and the kinetic (or radio jet) mode, which may dominate in different accretion states and at different cosmic epochs. In the case of kinetic-mode feedback, much of the energy output is released in mechanical form through collimated relativistic jets, which heat the surrounding gas and affect the large-scale environment on galaxy cluster scales. But such radio jet-driven feedback may be mostly limited to radio-loud sources, which only form a minority of the total AGN population. In contrast, radiation is the most direct and inevitable outcome of the accretion process, hence radiative-mode feedback can have a wider applicability. In addition, cosmic rays and magnetic fields may also contribute to the overall feedback process; for instance, it has been suggested that cosmic ray feedback should be included in simulations of galaxy formation \citep{Pfrommer_et_2017}. 

Powerful outflows are now commonly observed on galactic scales, with galactic outflows being often regarded as an empirical proof of AGN feedback in action \citep[][and references therein]{Fiore_et_2017, Fluetsch_et_2019}. 
However, the physical origin of the outflow driving mechanism is still much debated.
High-velocity winds can be driven by the AGN radiation field and be launched from the immediate vicinity of the central black hole. 
Such ultrafast outflows collide with the surrounding interstellar medium (ISM) and generate shockwaves propagating into the host galaxy \citep{Zubovas_King_2012, Faucher-Giguere_Quataert_2012}. Line-driven winds, powered by radiation pressure on spectral lines, have also been considered. Numerical simulations show that the local ultraviolet radiation from the accretion disc can launch line-driven disc winds from scales of a few hundred Schwarzschild radii \citep{Proga_et_2000}. Another possibility for driving large-scale feedback is via radiation pressure on dust, which can directly couple to the ISM on galactic scales \citep{Fabian_1999, Murray_et_2005, Thompson_et_2015}. Since the dust absorption cross section is much larger than the Thomson cross section ($\sigma_d \gg \sigma_T$), the radiation-matter coupling can be considerably enhanced. In the following, we thus focus on galactic-scale outflows driven by radiation pressure on dust.  

In the framework of the AGN radiative dusty feedback scenario, we have previously discussed how the observed dynamics and energetics of galactic outflows can be reproduced, provided that radiation trapping is properly taken into account \citep{Ishibashi_Fabian_2015, Ishibashi_et_2018}. 
Interestingly, recent radiation hydrodynamic (RHD) simulations of radiation pressure-driven shells indicate that the boost factor is roughly equal to the infrared optical depth (except for extreme optical depths), largely confirming the analytic picture \citep{Costa_et_2018a, Barnes_et_2018}. 
More recently, we have also considered the temporal evolution of the central AGN luminosity output [$L(t)$, with different forms of luminosity decay histories], and analysed its effects on the evolution of the radiation pressure-driven outflows.  
In particular, we discussed how the high observed values of the outflow energetics can be explained by either radiation trapping or  luminosity decay, and how the recently discovered `fossil' outflows may be interpreted as relics of past AGN activity \citep{Ishibashi_Fabian_2018}.

In most previous works, the central radiation was implicitly assumed to be isotropic, leading to the development of spherically symmetric outflows. However, in reality, the emission emerging from the accretion disc is intrinsically anisotropic, with the exact radiation pattern being determined by the central BH spin \citep[e.g.][]{Sun_Malkan_1989}. 
We recall that the last stable orbit is located at a smaller radius for a higher BH spin, whereby the radiative efficiency is higher, and the relativistic effects are stronger. 
We have recently discussed how such BH spin-induced radiation pattern may determine the geometry of the radiation pressure-driven outflows on galactic scales \citep{Ishibashi_et_2019}. 

Numerical simulations also suggest that the anisotropic radiation can have a dramatic effect on the outflow properties, but the anisotropy factor remains unconstrained and simply left as a free parameter \citep[e.g.][]{Williamson_et_2019}. 
A proper treatment of the underlying radiation field is thus crucial in order to correctly determine the shape and strength of galactic outflows driven by radiation pressure. 
While we focus here on the anisotropy of the radiation field, we note that anisotropic outflows may also develop as a result of non-spherical gas distributions in the host galaxy, such as in disc galaxies, and multi-dimensional effects should ideally be included  \citep{Roth_et_2012, Bieri_et_2017, Hartwig_et_2018, Barnes_et_2018, Menci_et_2019}. 

In this paper we investigate the angular dependence of the AGN luminosity output, and quantify its impact on the dynamics and energetics of the radiation pressure-driven outflows. In Sections $\ref{Sect_anisotropic_feedback}-\ref{Sect_obscuration}$, we analyse the important role of the directional dependence of the luminosity output, $L(\theta)$, determined by the BH spin-induced radiation pattern. In Section \ref{Sect_Dust}, we explore the impact of a spatially varying dust-to-gas ratio on the evolution of galactic outflows. 
We discuss the resulting physical implications on outflow propagation, obscuration geometry, and accretion process in Section \ref{Sect_discussion}.


\section{Anisotropic radiative feedback due to black hole spin}
\label{Sect_anisotropic_feedback}

\subsection{Radiative dusty feedback}

We consider AGN feedback driven by radiation pressure on dust, which sweeps up the ambient dusty gas into an outflowing shell.
The shell evolution is governed by the competition between the outward force due to radiation pressure and the inward force due to gravity, with the general form of the equation of motion given by (e.g. \citet{Thompson_et_2015, Ishibashi_Fabian_2015}): 
\begin{equation}
\frac{d}{dt} [M_{g}(r) v] = \frac{L(\theta)}{c} (1 + \tau_{IR} - e^{-\tau_{UV}} ) - \frac{G M(r) M_{g}(r)}{r^2} \, , 
\label{Eq_motion}
\end{equation} 
where $v$ is the shell velocity, $L(\theta)$ is the angle-dependent luminosity, $M(r)$ is the total mass distribution, $M_g(r)$ is the gas mass within radius $r$, $\tau_\mathrm{IR}$ and $\tau_\mathrm{UV}$ are the infrared (IR) and ultraviolet (UV) optical depths.   
We assume an isothermal potential, $M(r) = \frac{2 \sigma^2 r}{G}$, where $\sigma$ is the velocity dispersion; and we parametrise the 
ambient gas density distribution as a power-law of radius: $n(r) = n_0 (\frac{r}{R_0})^{-\alpha}$, where $\alpha$ is the power-law exponent, $n_0$ is the density of the external medium, and $R_0$ is the initial radius. 
We further assume an isothermal gas distribution, corresponding to the case $\alpha = 2$, such that the gas mass scales as $M_g(r) = 4 \pi m_p n_0 R_0^2 r$. Equivalently, the shell mass can also be expressed in terms of the gas fraction ($f_g$) as $M_g(r) = f_g M(r) = \frac{2 f_g \sigma^2}{G}r$. 

A major fraction of the AGN bolometric luminosity output is emitted in the `blue bump' component, peaking in the UV region (around energies of $E \sim 10$ eV). We consider here the UV luminosity, which is most efficiently absorbed by dust. UV photons are absorbed by dust grains embedded in the gas, and by energy conservation, they are reprocessed and re-emitted as IR photons ($E \sim 1$ eV). Dust opacities are greatest at UV wavelengths, and much lower in the IR band ($\kappa_\mathrm{UV} \gg \kappa_\mathrm{IR}$). The dust opacity is also lower at optical wavelengths than in the UV band, e.g. reduced by a factor of $\sim 4$ \citep{Hensley_et_2014}. Here we simply assume two energy bands, i.e. single-scattering UV and multiple-scattering IR, as in many previous studies \citep{Thompson_et_2015, Ishibashi_Fabian_2015}. We note that such a simplified two-band approach has also been adopted in numerical simulations \citep{Costa_et_2018b, Huang_et_2020}. In broad terms, the characteristic wavelengths around which the selected dust opacities may apply are $\lambda \sim 100$nm and $\lambda \sim 1 \mu m$ for the UV and IR bands, respectively. A typical value for the dust opacity at UV wavelengths is $\kappa_\mathrm{UV} \sim 10^3 cm^2/g$ for a Milky Way like dust-to-gas ratio \citep{Murray_et_2011, Heckman_Thompson_2019}. In the IR band, the Rosseland mean dust opacity can be roughly approximated by a constant value, $\kappa_\mathrm{IR} \sim 5 \, \mathrm{cm^2 g^{-1}}$, as the temperature dependence may be neglected for $T \gtrsim 150$ K \citep{Thompson_et_2015}. A more accurate treatment would require a multiband approach, comprising 5 or even 10 radiation bands, with separate opacities in each band \citep[e.g.][]{Hopkins_et_2020}. We will further discuss the issue of the wavelength-dependent dust opacities in Section \ref{Subsec_caveats_outlook}. 

In the following, we assume constant dust opacities, parametrised by $\kappa_\mathrm{UV} = 10^3 \, \mathrm{cm^2 g^{-1} f_{dg, MW}}$ and $\kappa_\mathrm{IR} = 5 \, \mathrm{cm^2 g^{-1} f_{dg, MW}}$, as adopted in previous works \citep{Thompson_et_2015, Ishibashi_Fabian_2015}. Note that the dust opacities directly scale with the dust-to-gas ratio ($f_{dg}$) normalised to the Milky Way value. The associated IR and UV optical depths are then given by $\tau_\mathrm{IR}(r) = (\kappa_\mathrm{IR} m_p n_0 R_0^2)/r$ and $\tau_\mathrm{UV}(r) = (\kappa_\mathrm{UV} m_p n_0 R_0^2)/r$, respectively. 
In terms of the gas fraction, the optical depths can also be equivalently written as $\tau_\mathrm{IR,UV} = \frac{\kappa_\mathrm{IR,UV} f_g \sigma^2}{2 \pi G r}$. 

From equation (\ref{Eq_motion}), the radiative force is given by: 
\begin{equation}
F_{rad} = \frac{L(\theta)}{c} (1 + \tau_{IR} - e^{-\tau_{UV}} ) \, , 
\label{Eq_Frad}
\end{equation}
where we explicitly consider the angular dependence of the luminosity output $L(\theta)$; while the gravitational force is given by: 
\begin{equation}
F_{grav} = \frac{G M(r) M_{g}(r)}{r^2} = 8 \pi m_p n_0 R_0^2 \sigma^2 \, . 
\label{Eq_Fgrav}
\end{equation}

By equating the outward force due to radiation pressure to the inward force due to gravity, a critical luminosity can be defined.  
In contrast to the case of isotropic radiation, there is not a one-single value of the critical luminosity, but rather an angle-dependent quantity. For instance, the condition $F_{rad} > F_{grav}$ may be satisfied along certain directions, but not in others. 


\subsection{BH spin-dependent radiation pattern}

The actual emission pattern emerging from the accretion disc is determined by the central BH spin. We recall that the dimensionless spin parameter ($a = cJ/GM^2$, where $M$ and $J$ are the mass and angular momentum of the BH) varies from $a = 0$ for a non-rotating Schwarzschild BH, to $a = 1$ for a maximally-rotating Kerr BH. 
The innermost stable circular orbit (ISCO) is located at $r = 6 r_g$ (where $r_g = GM/c^2$ is the gravitational radius) for a non-rotating Schwarzschild BH, with a corresponding radiative efficiency of $\epsilon \sim 0.06$; while the ISCO is located at $r = 1 r_g$ for a maximally-rotating Kerr BH, with a corresponding radiative efficiency of order $\epsilon \sim 0.4$ (prograde case). 

The exact radiation pattern is determined by the central BH spin, with the relativistic effects being more important around rapidly rotating black holes \citep{Cunningham_1975, Sun_Malkan_1989}. Here we compute the precise emission pattern by employing the KERRBB model implemented in {\small XSPEC} \citep{Li_et_2005}. KERRBB is a numerical code modelling a general relativistic accretion disc around a rotating Kerr black hole, including all relativistic effects such as Doppler boosting, gravitational redshift, and light bending. The code assumes a multitemperature blackbody model for the accretion disc, taking into account the disc self-irradiation, and also  comprises the limb-darkening effect. The model parameters include the BH mass, spin, accretion rate, and inclination angle. In our set-up, the BH mass and accretion rate are kept fixed, while we analyse variations in the BH spin and the angular dependence. 

Two key factors determine the emerging radiation pattern: for a rapidly rotating BH, the ISCO is located at a smaller radius, which implies a higher radiative efficiency as well as stronger relativistic effects. In fact, as the inner edge of the accretion disc moves closer to the centre, the photon trajectories can be significantly modified and distorted (e.g. due to light bending). 
Although the KERRBB code was originally developed to fit the spectra of stellar-mass BHs in X-ray binaries, the results can be applied to supermassive BHs, as we are only interested in the angular dependence of the radiation pattern. Indeed, by assuming scale-invariance, the use of the KERRBB model can be extended to describe the supermassive BH data and analytical approximations can  be derived \citep{Campitiello_et_2018}. 

\begin{figure}
\begin{center}
\includegraphics[angle=0,width=0.4\textwidth]{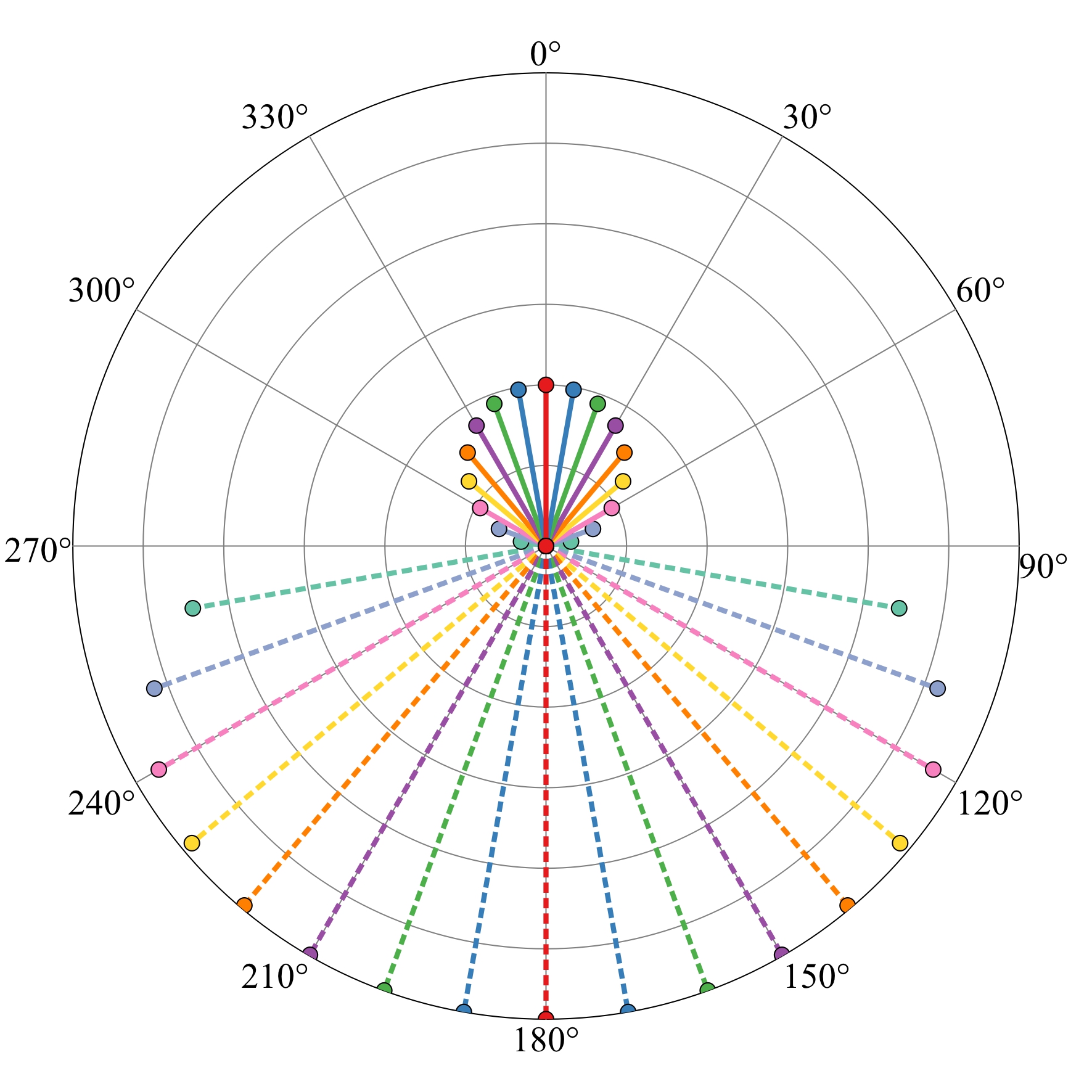}
\caption{ 
Polar diagram of the radiation pattern for zero spin ($a = 0$, upper half, solid lines) and maximum spin ($a = 0.998$, lower half, dotted lines), both including limb-darkening. The radiative flux values are normalised to the zero spin, pole-on value; with angle variations between $\theta = 0^{\circ}$ and $\theta = 80^{\circ}$ (every $10^{\circ}$). 
The radiation pattern is focused along the polar axis in the case of zero spin, while it is much more isotropic in the case of maximum spin. 
}
\label{polar_diagram}
\end{center}
\end{figure}

In Fig. \ref{polar_diagram}, we show the resulting polar diagram for the case of zero spin (`spin-zero' $a =0$, upper half) and maximum spin (`spin-max' $a = 0.998$, lower half), with the radiative fluxes normalised to the spin-zero, pole-on value. 
In the case of zero spin, the radiative flux is vertically focused along the polar axis, and steadily declines with increasing inclination angle; whereas in the case of maximum spin, the radiation pattern is much more isotropic. 
More precisely, in the case of zero spin, the radiative flux at $\theta = 85^{\circ}$ is reduced by a factor of $\sim 11.2$ compared to the pole-on value; while the corresponding reduction is only by a factor of $\sim 1.5$ in the case of maximum spin. 
We have previously discussed how low BH spins may give rise to polar/prolate outflows, while high BH spins may lead to quasi-spherical/oblate outflows; providing a potential way of constraining the BH spin from the observed morphology of galactic outflows \citep{Ishibashi_et_2019}. 

In addition to the angular dependence, we observe that the actual strength of the radiative output can be quite different: the radiative flux can be much higher for highly-spinning BHs, at all inclination angles. Comparing the two extreme cases (spin-max vs. spin-zero), the radiative flux is higher by a factor of $\sim 2.9$ at $\theta = 0^{\circ}$ and $\sim 21.7$ at $\theta = 85^{\circ}$. 
As a consequence, the difference between zero spin and maximum spin BHs is due to both their different radiative efficiencies and their respective angular emission patterns. 
We note that the total luminosity (integrated over solid angles) should satisfy the condition $\frac{1}{4 \pi} \int_{\Omega} L(\theta,a) = \epsilon(a) \dot{M}c^2$, where $\dot{M}$ is the mass accretion rate and $\epsilon(a)$ is the radiative efficiency that is determined by the BH spin. Thus the overall difference in the radiative output between spin-zero and spin-max objects should be given by the difference in their corresponding radiative efficiencies. Yet the important point here is that the BH spin does not only modify the location of the ISCO (hence the radiative efficiency), but also shapes the angular pattern of the emitted radiation through relativistic effects. 
The resulting differences will reflect on the propagation of the radiation pressure-driven outflows on galactic scales.


\section{Outflow dynamics and energetics}
\label{Sect_dynamics_energetics}

We now compute the evolution of the radiation pressure-driven outflows by explicitly taking into account the angular dependence of the luminosity output, $L(\theta)$, in the equation of motion (Eq. $\ref{Eq_motion}$). 
The following values are taken as fiducial parameters of the model: $L_0 = 5 \times 10^{46}$ erg/s, $n_0 \cong 3 \times 10^4 \mathrm{cm^{-3}}$ (or equivalently $f_g = 0.05$), $R_0 = 10$ pc, $\sigma = 200$ km/s, $\kappa_\mathrm{IR} = 5 \mathrm{cm^2/g}$, $\kappa_\mathrm{UV} = \mathrm{10^3 cm^2/g}$.
Assuming a standard $M-\sigma$ relation \citep{Kormendy_Ho_2013}, the black hole mass corresponding to $\sigma \sim 200$ km/s is $M \sim 3 \times 10^8 M_{\odot}$, with a corresponding Eddington luminosity of $\sim 4 \times 10^{46}$ erg/s, which is comparable to our fiducial luminosity $L_0$. 

\begin{figure}
\begin{center}
\includegraphics[angle=0,width=0.4\textwidth]{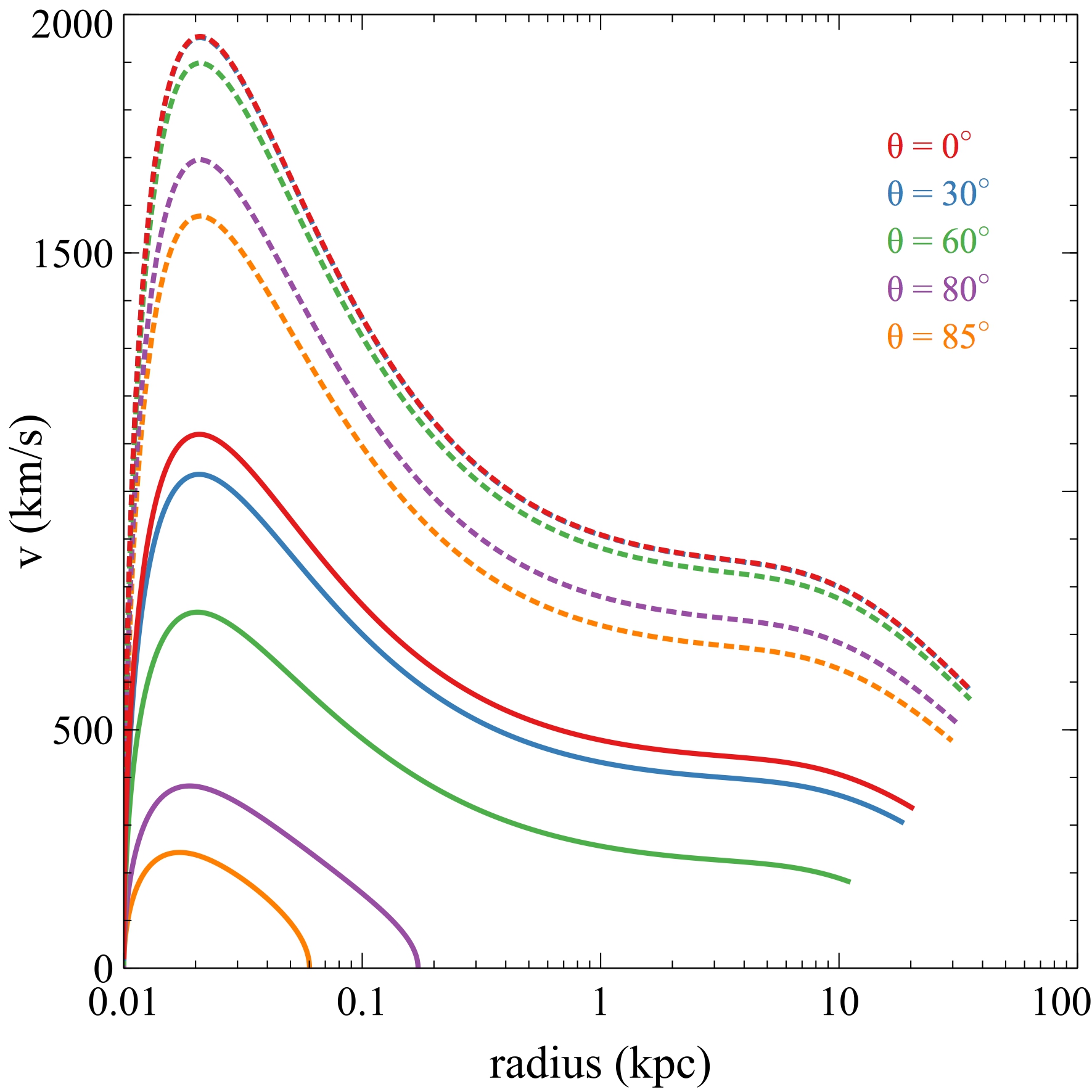}
\caption{ Radial velocity profiles of the outflowing shells in the case of zero spin (solid lines) and maximum spin (dotted lines), for different inclination angles: $\theta = 0^{\circ}$ (red), $\theta = 30^{\circ}$ (blue), $\theta = 60^{\circ}$ (green), $\theta = 80^{\circ}$ (violet), $\theta = 85^{\circ}$ (orange). 
Note that the blue dotted line ($\theta = 30^{\circ}$) is almost indistinguishable and overlaps with the red dotted line ($\theta = 0^{\circ}$). 
}
\label{plot_v_r}
\end{center}
\end{figure}

In Fig. $\ref{plot_v_r}$, we show the resulting radial velocity profiles of the outflowing shells in the case of zero spin (solid curves) and maximum spin (dotted curves), for different inclination angles. 
In the spin-zero configuration, and for low-to-moderate angles ($\theta \lesssim 60^{\circ}$), we see that the outflows can reach relatively large radii of order $r \gtrsim 10$ kpc, with velocities of a few hundred km/s. But for higher inclination angles ($\theta \gtrsim 80^{\circ}$), the outflowing shells only reach a maximal distance of $r \sim 0.1$ kpc, and eventually fall back. 
Hence the outcome can be completely different (large-scale propagation vs. fall-back), just due to the angular dependence. 
In contrast, in the case of maximum spin, the outflows reach large radii of tens of kpc, with speeds of several hundred km/s, at all inclination angles. The difference in the maximal distance reached (and the corresponding speed) is only by a factor of $\sim 1.2$ between $\theta = 0^{\circ}$ and $\theta = 85^{\circ}$. Hence the outflow behaviour is pretty similar for different inclination angles, indicative of a quasi-spherical propagation. 

In observational works, galactic outflows are often characterised by three physical quantities: the mass outflow rate ($\dot{M}$), the momentum flux ($\dot{p} = \dot{M} v$), and the kinetic power ($\dot{E}_{k} = \frac{1}{2} \dot{M} v^2$); as well as their two derived quantities: the momentum ratio ($\zeta = \frac{\dot{p}}{L/c}$) and the energy ratio ($\epsilon_k = \frac{\dot{E}_k}{L}$) \citep{Gonzalez-Alfonso_et_2017, Fiore_et_2017, Fluetsch_et_2019}. 
Bolometric luminosities are quoted in the observational works, while the difference with the driving UV luminosity may be at most of a factor of $\sim 2$, assuming a characteristic quasar spectrum \citep{Costa_et_2018b}. Other more important uncertainties are involved: most observational measurements report single global values of the outflow energetics at a given location and time. 
Future observations providing the radial and angular profiles of the outflow energetics will allow us more detailed comparisons with different model predictions \citep[see also][]{Costa_et_2018b, Hartwig_et_2018}. 

\begin{figure*}
\begin{multicols}{3}
    \includegraphics[width=\linewidth]{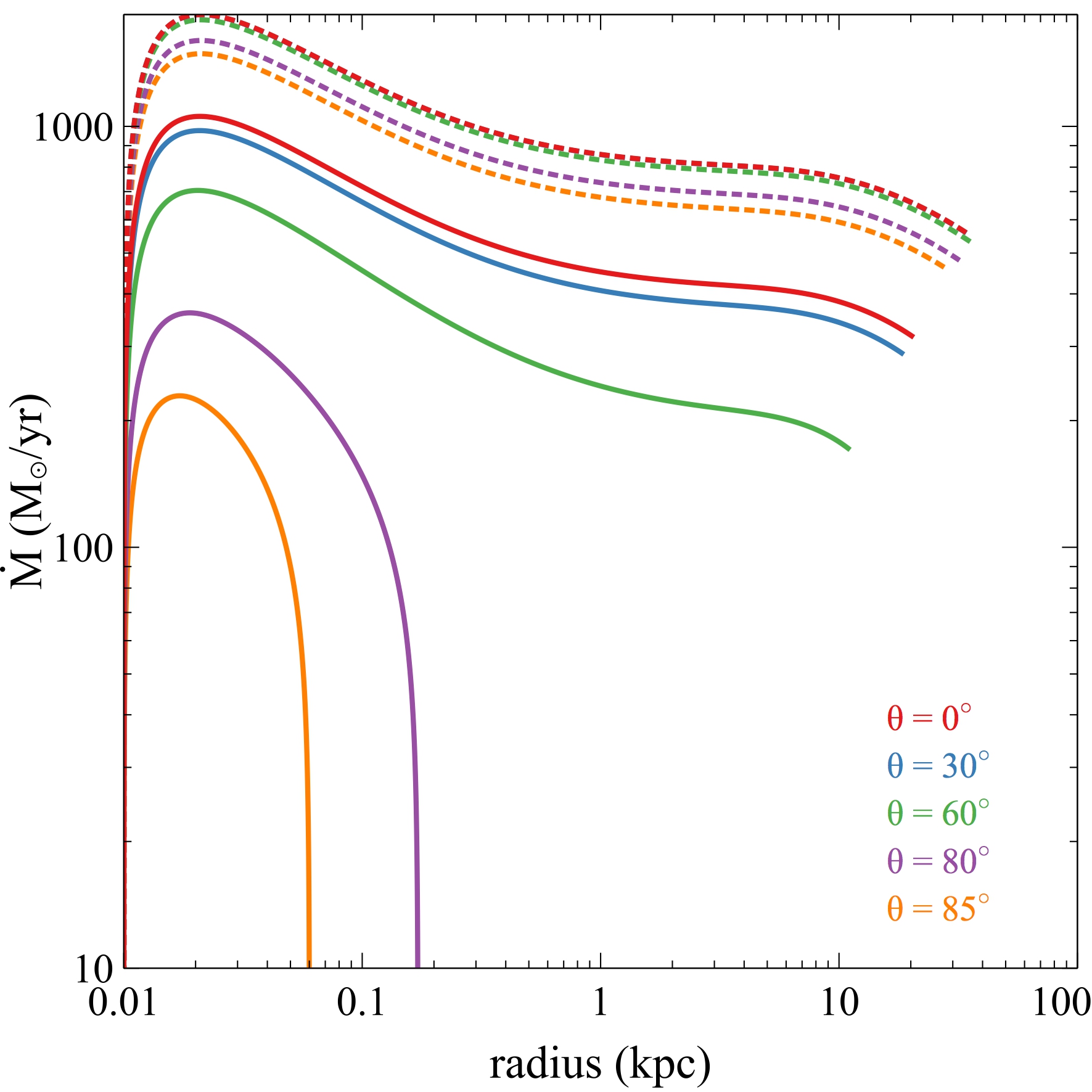}\par 
    \includegraphics[width=\linewidth]{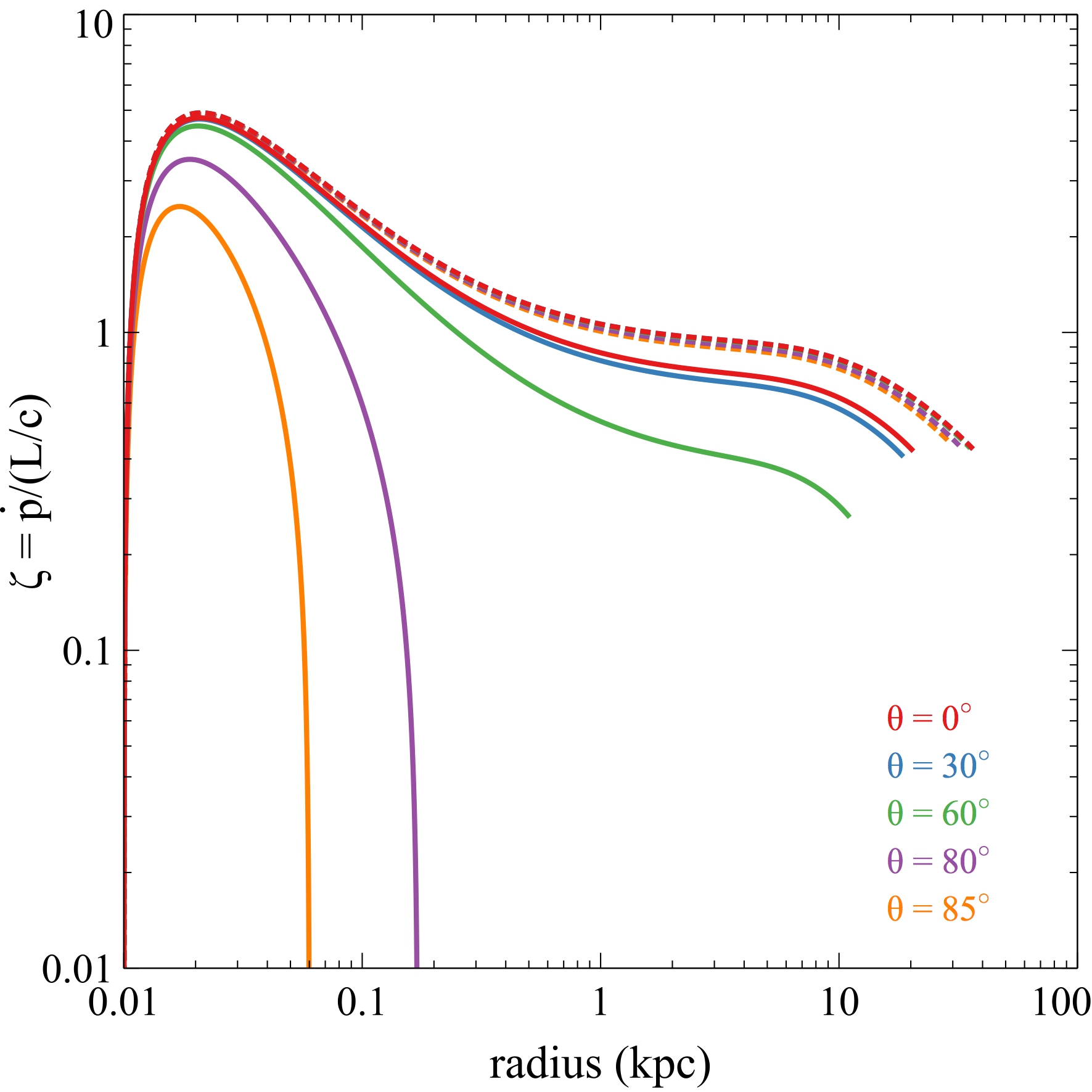}\par 
    \includegraphics[width=\linewidth]{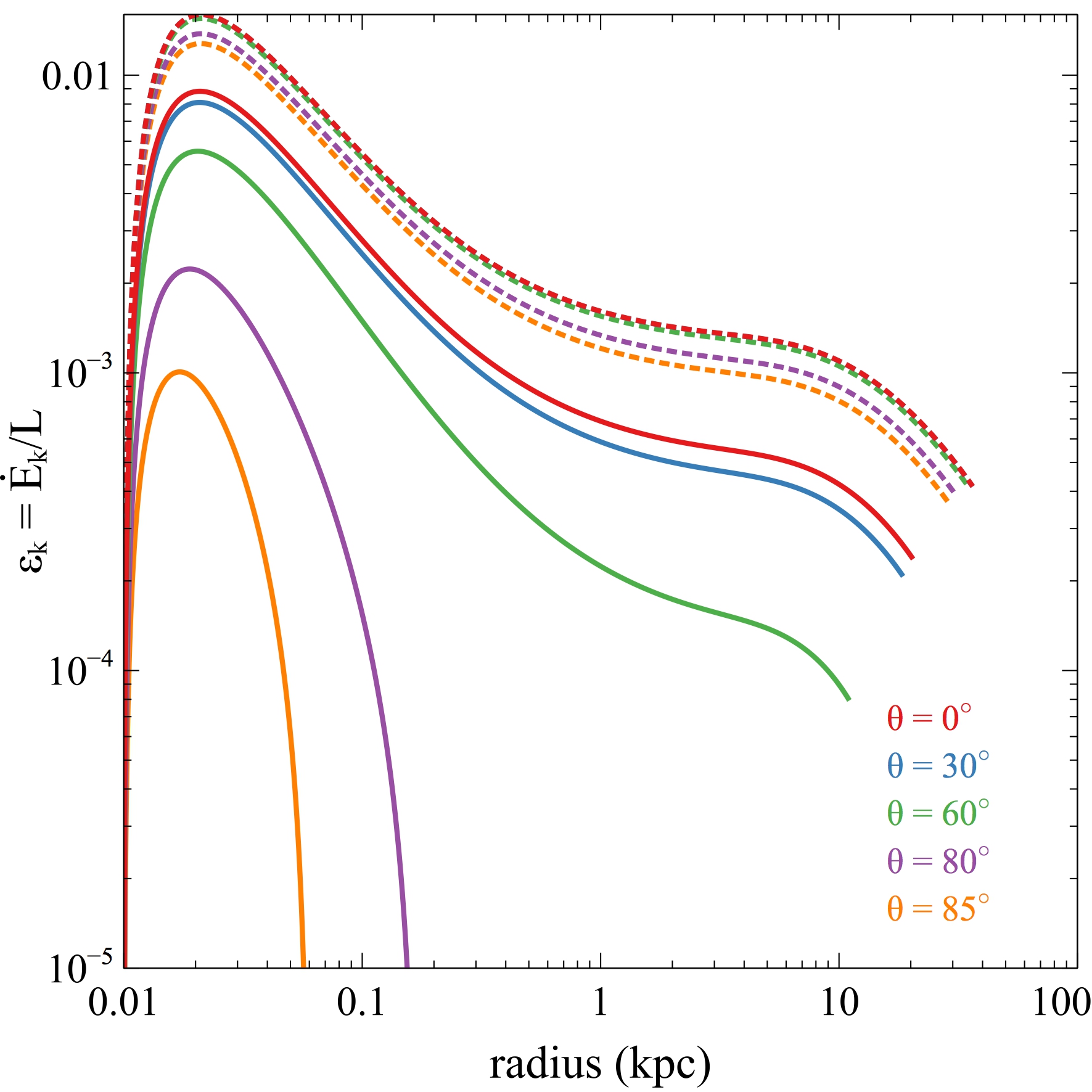}\par
    \end{multicols}
\caption{ 
Radial outflow energetics in the case of zero spin (solid lines) and maximum spin (dotted lines): mass outflow rate (left-hand panel), momentum ratio (middle panel), and energy ratio (right-hand panel). 
The colour coding for the different inclination angles is the same as in Fig. \ref{plot_v_r}: $\theta = 0^{\circ}$ (red), $\theta = 30^{\circ}$ (blue), $\theta = 60^{\circ}$ (green), $\theta = 80^{\circ}$ (violet), $\theta = 85^{\circ}$ (orange). 
Note that the blue dotted line ($\theta = 30^{\circ}$) is almost indistinguishable and overlaps with the red dotted line ($\theta = 0^{\circ}$).
}
\label{plot_energetics}
\end{figure*}

In Fig. $\ref{plot_energetics}$, we plot the radial profiles of the outflow energetics in the case of zero spin (solid curves) and maximum spin (dotted curves), for different inclination angles. In the case of zero spin, and for low-to-moderate angles ($\theta \lesssim 60^{\circ}$), the mass outflow rates reach values of several hundred $M_{\odot}/\mathrm{yr}$ on kpc-scales, while the momentum ratio and energy ratio are of the order of $\zeta \sim 1$ and $\epsilon_k \sim 10^{-3}$ on galactic scales. 
Observations of galactic outflows report typical momentum ratios in the range $\zeta \sim (1-30)$ and energy ratios in the range $\epsilon_k \sim (0.3 - 3) \%$ \citep{Fiore_et_2017, Fluetsch_et_2019}. The energetics of the radiation pressure-driven outflows at low angles are thus comparable to the observational values, especially matching the lower end of the observed range. 
[We have previously argued that radiation trapping is required in order to account for the highest values of the outflow energetics \citep{Ishibashi_et_2018}]. For higher inclination angles ($\theta \gtrsim 80^{\circ}$), the outflowing shells cannot propagate on large scales, and the associated energetics rapidly fall off, signalling some form of `failed' outflows. 
In contrast, in the case of maximum spin, the outflow energetics are quite similar for different inclination angles (with a maximal difference of the order of $\sim 1.3$ between $\theta = 0^{\circ}$ and $\theta = 85^{\circ}$). Thus outflows with reasonable energetics on kpc-scales can be driven at all inclination angles. Therefore rapidly-spinning BHs can drive powerful, quasi-spherical outflows on galactic scales; whereas slowly-rotating BHs can only drive weaker bipolar outflows in the polar regions (with failed outflows that tend to fall back near the equatorial plane). 

We have previously discussed how the temporal evolution of the radiative output, $L(t)$, can affect the outflow dynamics and energetics, for different forms of AGN luminosity decay \citep{Ishibashi_Fabian_2018}, or exponential growth in the early Universe \citep{Ishibashi_2019}. In the context of the wind feedback model, it has been argued that a power-law luminosity decay can adequately reproduce the observed range of momentum and energy loading factors \citep{Zubovas_2018}. We now realise that the angular dependence of the radiative output, $L(\theta)$, can be equally important in determining the outflow properties on galactic scales. 


\section{Obscuration}
\label{Sect_obscuration}

We next consider the obscuration properties associated with the propagation of dusty outflows driven by radiative feedback. 
For an isothermal gas distribution, the shell column density is given by
\begin{equation}
N_{sh}(r) = \frac{M_{g}(r)}{4 \pi m_p r^2} = \frac{n_0 R_0^2}{r} ,
\end{equation}

\begin{figure}
\begin{center}
\includegraphics[angle=0,width=0.4\textwidth]{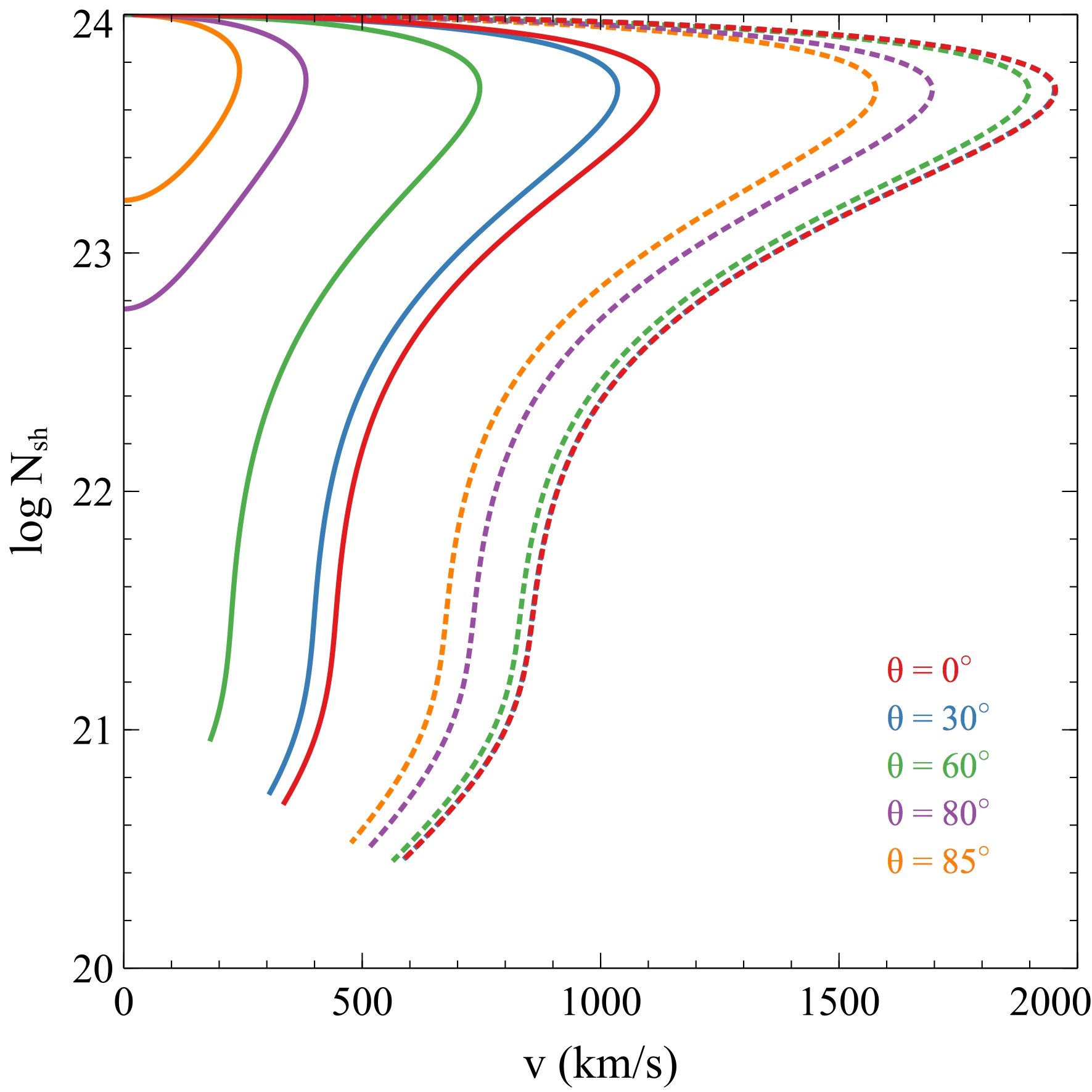} 
\caption{ 
Shell column density as a function of velocity for the case of zero spin (solid lines) and maximum spin (dotted lines). Same colour coding for the different inclination angles as in Fig. \ref{plot_v_r}: $\theta = 0^{\circ}$ (red), $\theta = 30^{\circ}$ (blue), $\theta = 60^{\circ}$ (green), $\theta = 80^{\circ}$ (violet), $\theta = 85^{\circ}$ (orange). 
Note that the blue dotted line ($\theta = 30^{\circ}$) is almost indistinguishable and overlaps with the red dotted line ($\theta = 0^{\circ}$).
}
\label{plot_N_v}
\end{center}
\end{figure} 

In Fig. $\ref{plot_N_v}$, we plot the shell column density as a function of velocity in the case of zero spin (solid curves) and maximum spin (dotted curves), again for different inclination angles. Starting from the same initial condition (corresponding to the Compton-thick limit $N_\mathrm{sh,0} \sim 10^{24} \mathrm{cm^{-2}}$), we observe that the subsequent evolution can be quite different: in the case of maximum spin, the shells are accelerated to high speeds and the associated column densities rapidly fall off; whereas in the case of zero spin, the column densities decrease much more slowly, suggesting that the obscuration is more long-lived. 
Within the spin-zero case, we also note significant differences due to the angular dependence: for low-to-moderate angles ($\theta \lesssim 60^{\circ}$), the outflows can propagate on large scales with relatively high speeds, effectively removing the obscuration in the polar directions; while for larger angles ($\theta \gtrsim 80^{\circ}$), the outflowing shells tend to fall back, leading to long-lived obscuration close to the equatorial plane. On the other hand, in the case of maximum spin, there is not much difference between the different angles, and the obscuration can be efficiently cleared out even at high inclination angles. 

Therefore highly-spinning BHs can efficiently remove dusty gas from almost all directions, via strong quasi-spherical outflows.
Since the obscuration is rapidly cleared out, the duration of the obscured phase should be rather short for high-spin objects. 
In contrast, slowly-rotating BHs are less efficient in clearing the obscuring columns, and can only remove gas from the polar regions; while the dusty gas survives in the equatorial directions, such that the corresponding obscured phase should be much more long-lived. 
A characteristic obscuration timescale (or blowout time) can be defined as the time required for the column density to drop to $\sim 10^{22} \mathrm{cm^{-2}}$, when the source becomes unobscured. Starting from the same initial conditions, a typical obscuration timescale may be of the order $\lesssim 10^6$ yr for a maximally rotating BH and a factor of $\sim 2$ longer for a zero spin object, in the polar direction ($\theta \sim 0^{\circ}$). The difference in the obscuration timescales between spin-zero and spin-max objects increases with increasing inclination angle, e.g. the difference may be $\sim (5-10)$ for $\theta \gtrsim 70^{\circ}$. 
In the case of zero-spin objects at higher inclination angles ($\theta \gtrsim 80^{\circ}$), the outflow is unable to propagate on large scales and tends to fall back, leading to long-lived obscuration near the equatorial plane.


\section{The impact of dust}
\label{Sect_Dust}

Up to now, we have implicitly assumed a constant dust-to-gas ratio ($f_{dg}$) throughout the outflow propagation in the host galaxy. This is clearly an optimistic assumption, since the dust content is unlikely to stay constant, due to the different dust destruction processes operating within a galaxy. In fact, the dust grains can be easily destroyed, e.g. via thermal/non-thermal sputtering in interstellar shock waves. 
Thus one should take into account dust destruction and the subsequent reduction in the amount of available dusty gas. 

\subsection{Spatially varying dust-to-gas ratio} 

In order to model dust depletion, we consider a radially declining dust-to-gas ratio as a power-law of radius with slope $\gamma$: 
\begin{equation}
f_{dg} = f_{dg,0} \left( \frac{r}{R_0} \right)^{-\gamma} \, , 
\end{equation}
where $f_{dg,0}$ is the initial dust-to-gas ratio at the initial radius $R_0$. 
Since the dust opacities scale with the dust-to-gas ratio, the effective optical depths will decrease accordingly with increasing distance from the centre. 

\begin{figure}
\begin{center}
\includegraphics[angle=0,width=0.4\textwidth]{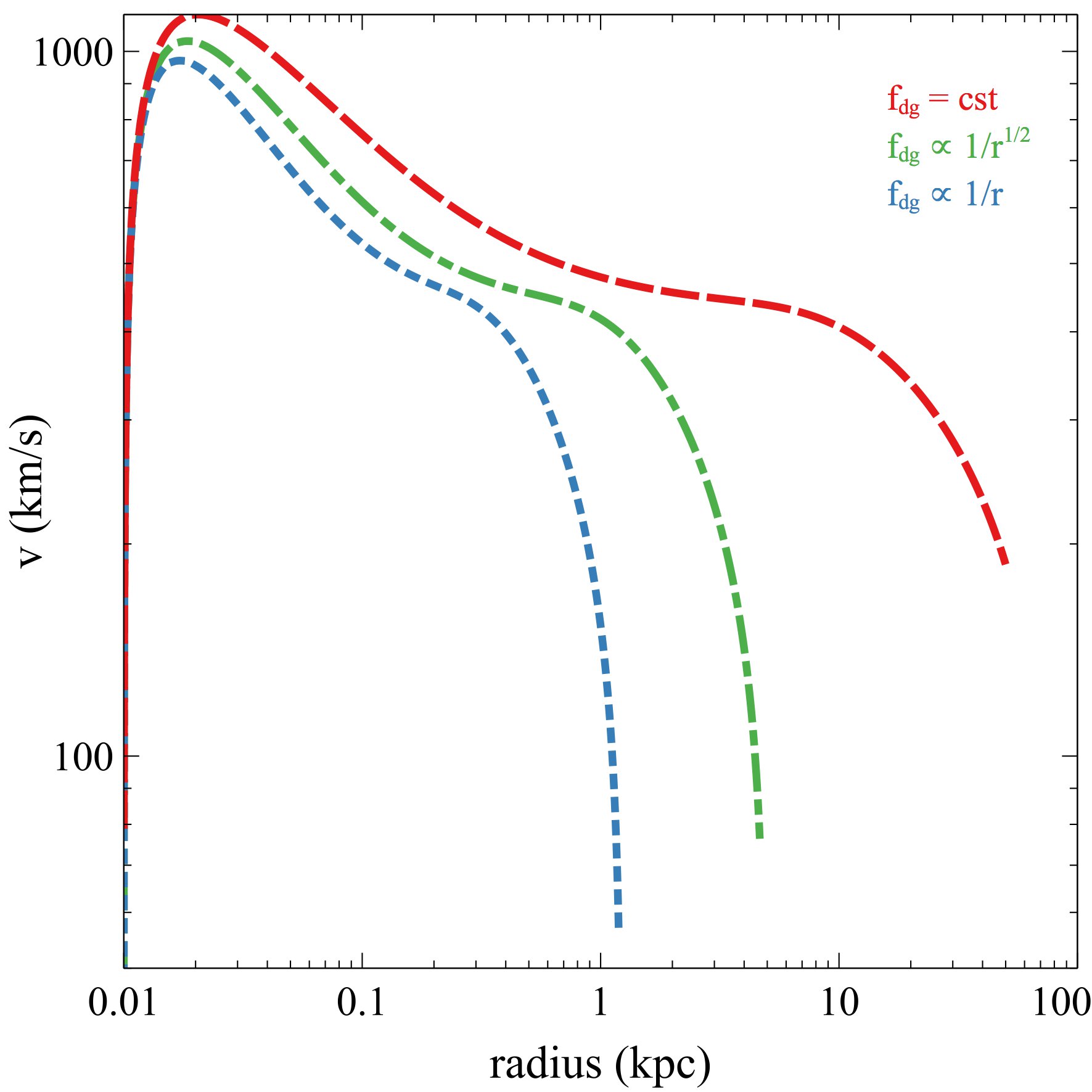}
\caption{ 
Radial velocity profiles for radially declining dust-to-gas ratios: $f_{dg} = \mathrm{cst}$ (red dashed), $f_{dg} \propto 1/r^{1/2}$ (green dash-dot), $f_{dg} \propto 1/r$ (blue dotted).  
}
\label{plot_v_r_varFdg_radial}
\end{center}
\end{figure} 

\begin{figure}
\begin{center}
\includegraphics[angle=0,width=0.4\textwidth]{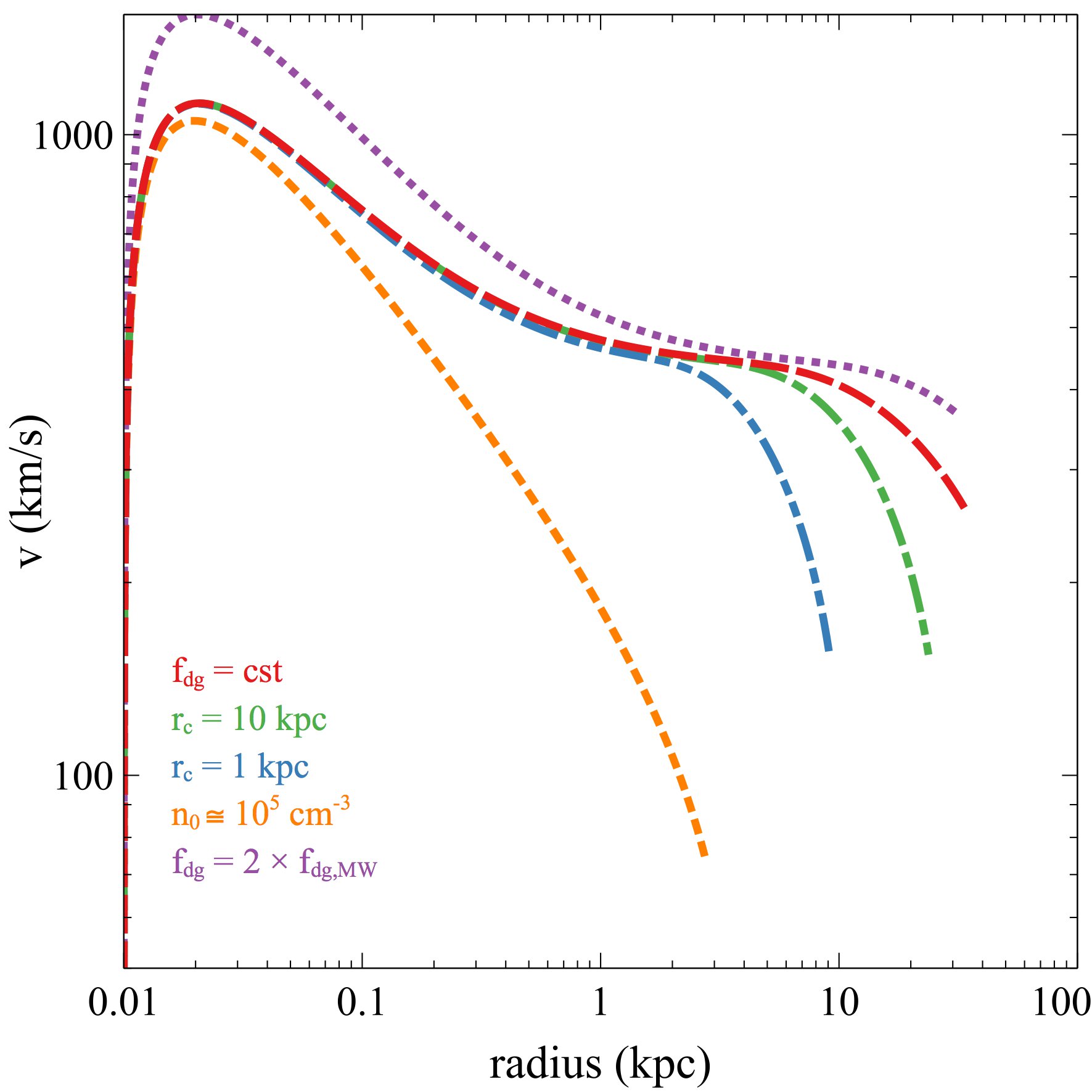}
\caption{ 
Radial velocity profiles for variations in the dust-to-gas ratio: $f_{dg} = \mathrm{cst}$ (red dashed), radially declining $f_{dg}$ with $r_c = 10$ kpc (green dash-dot) and with $r_c = 1$ kpc (blue dash-dot-dot). 
Additional cases with enhanced external density $n_0 \cong 10^5 \mathrm{cm^{-3}}$, corresponding to gas fraction $f \sim 0.2$ (orange dotted), and enhanced dust-to-gas ratio $f_{dg} = 2 \times f_\mathrm{dg,MW}$ (violet fine-dotted). 
}
\label{plot_v_r_varFdg_variations}
\end{center}
\end{figure}

Figure $\ref{plot_v_r_varFdg_radial}$ shows the effect of a radially declining dust-to-gas ratio on the propagation of the outflowing shell. The radial velocity is plotted for three different values of the  power-law slope: $\gamma = 0,1,2$.
For a constant dust-to-gas ratio (i.e. $\gamma = 0$), the outflowing shell can reach large radii ($r > 10$ kpc); whereas for a radially declining dust-to-gas ratio with $\gamma = 1$ ($f_{dg} \propto 1/r$), the shell turns around and starts to fall back after reaching a maximal distance of $r \sim 1$ kpc.  We see that for the less steeply declining case with $\gamma = 1/2$ ($f_{dg} \propto 1/r^{1/2}$), the outflowing shell presents an intermediate behaviour. Therefore a spatially varying dust-to-gas ratio can have a significant impact on the propagation of radiation pressure-driven outflows. 

On the other hand, the dust content may not monotonically decrease with increasing distance from the centre, as the dust is not only destroyed on galactic scales. In fact, dust grains can also be created in supernova explosions and released in the surrounding environment. Indeed, recent observations indicate that substantial quantities of dust can be produced in core-collapse supernovae \citep{Owen_Barlow_2015, Wesson_et_2015}. Furthermore, grain growth in the interstellar medium can also contribute to the dust mass increase \citep{Michalowski_2015}, with turbulence further accelerating the growth of dust grains \citep{Mattsson_2020}. 
As a consequence, significant amounts of dust can still be present on galactic scales, with the dust content only starting to decline beyond a certain critical radius.

We next model the radially declining dust profile as $f_{dg} \propto \frac{1}{r(r+r_c)}$, where $r_c$ is a characteristic radius. In Figure $\ref{plot_v_r_varFdg_variations}$, we show the resulting effect for two different values of $r_c = 1$ kpc and $r_c = 10$ kpc. We observe that the outflow evolution is roughly the same as for the case of a constant dust-to-gas ratio up to the characteristic radius $r_c$, and only starts to diverge at larger radii. A similar behaviour was also observed by \citet{Costa_et_2018a} who mimicked dust destruction by assuming that the swept-up mass becomes dust-free beyond a certain critical radius.

Figure $\ref{plot_v_r_varFdg_variations}$ also shows the effects of varying the value of the dust-to-gas ratio and the external density (or equivalently gas fraction). For an enhanced dust-to-gas ratio ($f_{dg} = 2 \times f_\mathrm{dg,MW}$), the outflow reaches higher velocities and larger radii; while for an increased gas fraction ($f_g = 0.2$), the shell propagation is hindered due to the larger amount of swept-up material. The latter trend may be explained by considering the competition between the radiative and gravitational forces, which governs the outflow dynamics. From equations (\ref{Eq_Frad}) and (\ref{Eq_Fgrav}), we recall that the radiative force is given by $F_{rad} = \frac{L(\theta)}{c} (1 + \tau_{IR} - e^{-\tau_{UV}} )$, where $\tau_\mathrm{IR,UV} = \frac{\kappa_\mathrm{IR,UV} f_g \sigma^2}{2 \pi G r}$, while the gravitational force is given by $F_{grav} = \frac{G M(r) M_{g}(r)}{r^2} = \frac{4 f_g \sigma^4}{G}$. We note that the gravitational force is a constant independent of radius, while it scales with the gas fraction. On the other hand, the IR optical depth also scales with the gas fraction, but decreases with radius as $\propto 1/r$. Therefore, although the IR optical depth increases with increasing gas fraction, the radiative force in the IR-optically thick regime falls off with radius; whereas the gravitational force (proportional to $f_g$) remains constant. A more detailed analysis of the dependence of the outflow dynamics and energetics on the different underlying parameters (such as $L$, $M_{sh}$, and $f_{dg}$) in different optical depth regimes has been previously discussed \citep{Ishibashi_et_2018, Ishibashi_Fabian_2018}. 


\section{Discussion}
\label{Sect_discussion}

\subsection{Anisotropic outflows and obscuration}

In realistic situations, we need to take into account the anisotropic nature of the accretion disc emission, which is set by the central BH spin. As we have seen, the resulting angular dependence of AGN radiative feedback can have a considerable impact on the dynamics and energetics of the radiation pressure-driven outflows (Section \ref{Sect_dynamics_energetics}).
Actually, 3D radiation hydrodynamic simulations indicate that the anisotropy of the radiation field can be equally important than the Eddington ratio in determining the outflow properties \citep{Williamson_et_2019}. This underlines the importance of properly constraining the anisotropy factor (which is left as a free parameter in the simulations). 
Here we exactly quantify the anisotropy, based on the physically motivated radiation pattern obtained with the KERRBB code, and we analyse the quantitative effects on the outflow properties.

Since the actual radiation pattern is determined by the BH spin, we obtain significant differences in the outflow dynamics and energetics, for low-spin and high-spin sources. In the case of maximum spin, powerful outflows can propagate on large scales at all inclination angles, with uniform energetics ($\zeta \sim 1$ and $\epsilon_k \gtrsim 10^{-3}$), comparable to values observed in galactic outflows \citep{Fluetsch_et_2019}. Instead, in the case of zero spin, the outflows can only propagate with reasonable energetics in the polar directions. We may then expect the development of wide-angle bipolar outflows, with modest energetics, for the majority of low-to-moderate BH spins.  

From the observational perspective, there is growing evidence of polar dust emission, detected on scales of $\sim (1-100)$ pc, in a number of local AGNs \citep[e.g.][and references therein]{Asmus_2019}. 
Such polar dust emission is not expected in classical AGN unification scenarios \citep{Antonucci_1993}, but may originate in polar dusty winds driven by radiation pressure. Recent analysis of VLTI/MIDI data suggests a correlation between the amount of dust ejected in the wind and the Eddington ratio, supporting the notion that the dusty winds are indeed driven by radiation pressure \citep{Leftley_et_2019}. 
  
The combined effects of the angular dependence and radiative output (both dictated by the BH spin) directly affect the AGN outflow and obscuration properties (Sections \ref{Sect_dynamics_energetics} and \ref{Sect_obscuration}). As a consequence, we predict two distinct trends for high-spin and low-spin objects. 
Highly-spinning BHs have high radiative output and drive strong quasi-spherical outflows, which clear out the dusty gas from almost all directions. Since the obscuration is rapidly and efficiently removed, these intrinsically bright sources are also more likely to be unobscured, hence preferentially detected in observational samples. In fact, this can lead to severe selections biases, especially in flux-limited AGN surveys \citep{Vasudevan_et_2016}. 
In contrast, slowly-rotating BHs have lower radiative output, and can only drive weaker bipolar outflows that remove gas from polar angles, but much of the obscuring gas may stay behind in the equatorial regions. Such intrinsically dimmer sources are also more likely to remain obscured, and can be easily missed in observational surveys. 
This could be related to the fact that the bulk of the AGN population seems to be obscured, with the most heavily obscured sources currently eluding detection \citep[][and references therein]{Hickox_Alexander_2018}. 


\subsection{Connection to accretion}

We recall that the obscuring material also forms part of the accreting matter feeding the central BH. 
Assuming Eddington-limited accretion, the BH mass grows in time as
\begin{equation} 
M(t) = M_0 \exp \left( \frac{1-\epsilon}{\epsilon} \frac{t}{t_E} \right) \, ,
\label{Eq_BHgrowth}
\end{equation} 
where $M_0$ is the initial BH seed mass, $\epsilon$ is the radiative efficiency, and $t_E = \sigma_T c/4 \pi G m_p$ is the Eddington timescale. 
Higher BH spins imply higher radiative efficiencies, thus the mass growth can be severely limited (as can be seen from equation \ref{Eq_BHgrowth}); hence rapidly-rotating BHs cannot grow efficiently. 
In fact, low BH spins with correspondingly low radiative efficiencies, seem to be required in order to allow rapid BH growth, favouring some form of `chaotic accretion' \citep{King_Pringle_2006}. 
Since low BH spins are less effective in generating powerful AGN feedback, it has been argued that low-spin BHs should be the most massive ones \citep{Zubovas_King_2019}.

Here we introduce another dimension, namely the BH spin-induced angular dependence.  
As we have seen, a high BH spin can drive quasi-spherical outflows, which can remove gas and thus prevent accretion, from most directions (except in the equatorial plane). In contrast, a low BH spin can only drive bipolar outflows, which remove gas from the polar regions, but accretion can still proceed from a wider range of directions (i.e. not confined within the equatorial plane). 

Radiation hydrodynamic simulations suggest that under anisotropic radiative feedback, the accretion rate can be considerably enhanced, thus allowing a rapid BH growth at early times \citep{Sugimura_et_2017}. 
In fact, numerical simulations suggest that the radiative flux in the equatorial plane is not strong enough, such that the inflowing gas can accrete relatively unimpeded by radiative feedback; while in the polar directions the radiative output may be viewed as super-Eddington \citep{Takeo_et_2018}. Rapid and efficient accretion, enabled by anisotropic radiative feedback, can have important implications for the growth of black holes in the early Universe. In particular, if the growth mode is extremely efficient, Population III star remnants could be envisaged as viable supermassive BH seeds \citep{Sugimura_et_2017}. 
However, the origin of the anisotropic radiative feedback, which ultimately enables rapid BH growth, is not physically motivated and remains uncertain; in particular, no explicit connection is made with the BH spin. 

In our picture, the nature of anisotropic feedback and its effects on BH growth can be directly interpreted in terms of the BH spin. Indeed, the central BH spin has two macroscopic effects: the strength of the radiative output and its directional dependence, which both combine to facilitate/hinder accretion for low/high spin BHs, thus governing the overall accretion history. 
High-spin BHs can drive powerful quasi-spherical outflows that remove accreting gas from most directions, hindering efficient accretion and BH growth. As a result, rapidly-spinning objects should mostly have lower BH masses (unless large masses are attained by other means, e.g. via mergers). Conversely, low-spin BHs drive weaker bipolar outflows, which allow efficient accretion and thus rapid BH growth. As a consequence, slowly-rotating BHs should be more massive, as they can acquire larger masses due to weaker anisotropic feedback. This is somewhat similar to the conclusion that the slowest spinning BHs should be the most massive ones \citep{Zubovas_King_2019}, but with the addition of the angular dimension. This may also be qualitatively consistent with the observational trend suggesting that the more massive BHs tend to have lower BH spins \citep[][and references therein]{Reynolds_2019}.

We further note that the BH spin is often invoked in the context of relativistic jets, with high spin values required to drive powerful radio jets (e.g. the Blandford-Znajek mechanism and variants thereof). However, here we find that the BH spin can actually influence the galactic outflows, solely due to radiative feedback, irrespective of the presence of radio jets. 


\subsection{Caveats and outlook}
\label{Subsec_caveats_outlook}

We have explored the impact of the anisotropic radiation on outflow propagation, AGN obscuration, and BH accretion. Here we focus on the anisotropy of the radiation field, shaped by the BH spin, while we assume a spherically symmetric and homogeneous gas distribution. Such a quasi-spherical configuration may be viable for spheroidal galaxies, and most relevant for the early enshrouded phase of buried quasars and ULIRG-like systems. In the case of smooth elliptical galaxies, the outflow geometry may then simply follow the underlying radiation pattern. But this may not be the case for disc/spiral galaxies, in which the outflow geometry is more likely shaped by the large-scale galactic disc.  

Indeed, another cause of anisotropy is the non-spherical distribution of the ambient gas in the host galaxy. Considering a disc geometry for the gas density profile, \citet{Hartwig_et_2018} and \citet{Menci_et_2019} study the 2D effect on the outflow evolution in the framework of the wind-shock feedback model. They find that galactic outflows tend to propagate along paths of least resistance, i.e. in the direction perpendicular to the galactic disc. Significant differences in the outflow dynamics and energetics are reported between the 2D and 1D models, depending on the details of the radiation-matter coupling.  

This is also closely related to the question of radiation trapping and photon leakage in realistic environments. The issue has been investigated by means of different numerical simulations, with somewhat contrasting results reported in the literature \citep[e.g.][]{Krumholz_Thompson_2013, Zhang_Davis_2017}. RHD simulations of outflows in disc galaxies with inhomogeneous interstellar medium indicate that the photons tend to escape through lower density channels, such that the number of IR multiscattering events is roughly one quarter of the IR optical depth \citep{Bieri_et_2017}. 3D radiative transfer calculations, including multidimensional effects,  show that the radiation force on dusty gas may be reduced by a factor of $\sim 2$ in cases of severe clumping compared to the case of a smooth gas distribution \citep{Roth_et_2012}. Recent RHD simulations of radiation pressure-driven outflows indicate that the boost factor is roughly equal to the IR optical depth, unless the latter is extremely high ($\tau_\mathrm{IR} \sim 100$) such that the diffusion time becomes longer than the flow time \citep{Costa_et_2018a}.

Although the question is not definitively settled yet, there seems to be a broad agreement on the relative importance of radiation trapping in the inner regions and at early times. 
In fact, AGN radiative feedback due to partial photon trapping must still play a crucial role in at least initiating the outflow. In future studies, we may wish to extend our model of radiative dusty feedback to the case of non-spherical gas geometries (e.g. disc galaxies) and more realistic gas density distributions (e.g. including a clumpy ISM). A thorough analysis of the interplay between the BH spin-induced anisotropic radiation field and the non-spherical and inhomogeneous gas distribution in the host galaxy should allow us to better characterise the role of AGN radiative feedback. 

In addition, one should also consider other important effects related to the AGN spectral energy distribution and the dust physics that are not considered here. For instance, it is well known that the peak emission of the accretion disc spectrum scales with the black hole mass, with the peak frequency ($\nu_p$) shifting towards lower values for larger masses (roughly scaling as $\nu_p \propto M^{-1/4}$, for a given Eddington ratio). Thus one may expect a certain BH mass dependence, which may influence the overall radiative output. 
Another important concern is the wavelength-dependence of the dust opacities. For simplicity, here we assume effective opacities, with constant representative values in the IR and UV bands, as also adopted in several other works \citep{Thompson_et_2015, Costa_et_2018b, Huang_et_2020}. 
In reality, the wavelength-dependence of the dust opacity should be properly included; the actual dust opacity depends on grain size, structure, and composition, as well as incident radiation spectrum. For instance, the dust opacity decreases with increasing wavelength in the IR band: e.g. assuming that the dust is mainly composed of silicate, the opacity almost steadily declines for $\lambda \gtrsim 10 \mu m$ \citep{Sarangi_et_2019}. Due to the reduced IR opacities at longer wavelengths, the effect of multiscattering could be somewhat overestimated in our analytic modelling. On the other hand, coagulation and grain growth form larger aggregates, which are thought to be responsible for the increase in the dust opacity observed at long wavelengths in the far-IR \citep[e.g.][and references therein]{Ysard_et_2018}.
In order to better constrain the impact of the wavelength-dependent dust opacities, as well as the black hole mass dependence, we will need to perform more detailed radiative transfer calculations (e.g. {\small CLOUDY}). 

Despite some limitations, we quantify here the general statement that high BH spins produce stronger feedback, and hence limit accretion, both in terms of outflow strength and directional dependence. We conclude that the BH spin-induced anisotropic feedback may play a major role in shaping AGN evolution over cosmic time.
 

\section*{Acknowledgements }

WI acknowledges support from the University of Zurich.

  
\bibliographystyle{mn2e}
\bibliography{biblio.bib}

\begin{thebibliography}{}

\bibitem[\protect\citeauthoryear{{Antonucci}}{{Antonucci}}{1993}]{Antonucci_1993}
{Antonucci} R.,  1993, \araa, 31, 473

\bibitem[\protect\citeauthoryear{{Asmus}}{{Asmus}}{2019}]{Asmus_2019}
{Asmus} D.,  2019, \mnras, 489, 2177

\bibitem[\protect\citeauthoryear{{Barnes}, {Kannan}, {Vogelsberger} \&
  {Marinacci}}{{Barnes} et~al.}{2018}]{Barnes_et_2018}
{Barnes} D.~J.,  {Kannan} R.,  {Vogelsberger} M.,    {Marinacci} F.,  2018,
  arXiv e-prints, p. arXiv:1812.01611

\bibitem[\protect\citeauthoryear{{Bieri}, {Dubois}, {Rosdahl}, {Wagner}, {Silk}
  \& {Mamon}}{{Bieri} et~al.}{2017}]{Bieri_et_2017}
{Bieri} R.,  {Dubois} Y.,  {Rosdahl} J.,  {Wagner} A.,  {Silk} J.,    {Mamon}
  G.~A.,  2017, \mnras, 464, 1854

\bibitem[\protect\citeauthoryear{{Campitiello}, {Ghisellini}, {Sbarrato} \&
  {Calderone}}{{Campitiello} et~al.}{2018}]{Campitiello_et_2018}
{Campitiello} S.,  {Ghisellini} G.,  {Sbarrato} T.,    {Calderone} G.,  2018,
  \aap, 612, A59

\bibitem[\protect\citeauthoryear{{Costa}, {Rosdahl}, {Sijacki} \&
  {Haehnelt}}{{Costa} et~al.}{2018a}]{Costa_et_2018a}
{Costa} T.,  {Rosdahl} J.,  {Sijacki} D.,    {Haehnelt} M.~G.,  2018a, \mnras,
  473, 4197

\bibitem[\protect\citeauthoryear{{Costa}, {Rosdahl}, {Sijacki} \&
  {Haehnelt}}{{Costa} et~al.}{2018b}]{Costa_et_2018b}
{Costa} T.,  {Rosdahl} J.,  {Sijacki} D.,    {Haehnelt} M.~G.,  2018b, \mnras,
  479, 2079

\bibitem[\protect\citeauthoryear{{Cunningham}}{{Cunningham}}{1975}]{Cunningham_1975}
{Cunningham} C.~T.,  1975, \apj, 202, 788

\bibitem[\protect\citeauthoryear{{Fabian}}{{Fabian}}{1999}]{Fabian_1999}
{Fabian} A.~C.,  1999, \mnras, 308, L39

\bibitem[\protect\citeauthoryear{{Fabian}}{{Fabian}}{2012}]{Fabian_2012}
{Fabian} A.~C.,  2012, \araa, 50, 455

\bibitem[\protect\citeauthoryear{{Faucher-Gigu{\`e}re} \&
  {Quataert}}{{Faucher-Gigu{\`e}re} \&
  {Quataert}}{2012}]{Faucher-Giguere_Quataert_2012}
{Faucher-Gigu{\`e}re} C.-A.,  {Quataert} E.,  2012, \mnras, 425, 605

\bibitem[\protect\citeauthoryear{{Fiore}, {Feruglio}, {Shankar}, {Bischetti},
  {Bongiorno}, {Brusa}, {Carniani}, {Cicone}, {Duras}, {Lamastra}, {Mainieri},
  {Marconi}, {Menci}, {Maiolino}, {Piconcelli}, {Vietri} \&
  {Zappacosta}}{{Fiore} et~al.}{2017}]{Fiore_et_2017}
{Fiore} F.,  {Feruglio} C.,  {Shankar} F.,  {Bischetti} M.,  {Bongiorno} A.,
  {Brusa} M.,  {Carniani} S.,  {Cicone} C.,  {Duras} F.,  {Lamastra} A.,
  {Mainieri} V.,  {Marconi} A.,  {Menci} N.,  {Maiolino} R.,  {Piconcelli} E.,
  {Vietri} G.,    {Zappacosta} L.,  2017, \aap, 601, A143

\bibitem[\protect\citeauthoryear{{Fluetsch}, {Maiolino}, {Carniani}, {Marconi},
  {Cicone}, {Bourne}, {Costa}, {Fabian}, {Ishibashi} \& {Venturi}}{{Fluetsch}
  et~al.}{2019}]{Fluetsch_et_2019}
{Fluetsch} A.,  {Maiolino} R.,  {Carniani} S.,  {Marconi} A.,  {Cicone} C.,
  {Bourne} M.~A.,  {Costa} T.,  {Fabian} A.~C.,  {Ishibashi} W.,    {Venturi}
  G.,  2019, \mnras, 483, 4586

\bibitem[\protect\citeauthoryear{{Gonz{\'a}lez-Alfonso}, {Fischer}, {Spoon},
  {Stewart}, {Ashby}, {Veilleux}, {Smith}, {Sturm} \& et
  al.}{{Gonz{\'a}lez-Alfonso} et~al.}{2017}]{Gonzalez-Alfonso_et_2017}
{Gonz{\'a}lez-Alfonso} E.,  {Fischer} J.,  {Spoon} H.~W.~W.,  {Stewart} K.~P.,
  {Ashby} M.~L.~N.,  {Veilleux} S.,  {Smith} H.~A.,  {Sturm} E.,    et al.
  2017, \apj, 836, 11

\bibitem[\protect\citeauthoryear{{Hartwig}, {Volonteri} \& {Dashyan}}{{Hartwig}
  et~al.}{2018}]{Hartwig_et_2018}
{Hartwig} T.,  {Volonteri} M.,    {Dashyan} G.,  2018, \mnras, 476, 2288

\bibitem[\protect\citeauthoryear{{Heckman} \& {Thompson}}{{Heckman} \&
  {Thompson}}{2017}]{Heckman_Thompson_2019}
{Heckman} T.~M.,  {Thompson} T.~A.,  2017, arXiv e-prints, p. arXiv:1701.09062

\bibitem[\protect\citeauthoryear{{Hensley}, {Ostriker} \& {Ciotti}}{{Hensley}
  et~al.}{2014}]{Hensley_et_2014}
{Hensley} B.~S.,  {Ostriker} J.~P.,    {Ciotti} L.,  2014, \apj, 789, 78

\bibitem[\protect\citeauthoryear{{Hickox} \& {Alexander}}{{Hickox} \&
  {Alexander}}{2018}]{Hickox_Alexander_2018}
{Hickox} R.~C.,  {Alexander} D.~M.,  2018, \araa, 56, 625

\bibitem[\protect\citeauthoryear{{Hopkins}, {Grudi{\'c}}, {Wetzel},
  {Kere{\v{s}}}, {Faucher-Gigu{\`e}re}, {Ma}, {Murray} \& {Butcher}}{{Hopkins}
  et~al.}{2020}]{Hopkins_et_2020}
{Hopkins} P.~F.,  {Grudi{\'c}} M.~Y.,  {Wetzel} A.,  {Kere{\v{s}}} D.,
  {Faucher-Gigu{\`e}re} C.-A.,  {Ma} X.,  {Murray} N.,    {Butcher} N.,  2020,
  \mnras, 491, 3702

\bibitem[\protect\citeauthoryear{{Huang}, {Davis} \& {Zhang}}{{Huang}
  et~al.}{2020}]{Huang_et_2020}
{Huang} X.,  {Davis} S.~W.,    {Zhang} D.,  2020, \apj, 893, 50

\bibitem[\protect\citeauthoryear{{Ishibashi}}{{Ishibashi}}{2019}]{Ishibashi_2019}
{Ishibashi} W.,  2019, \mnras, 489, 5225

\bibitem[\protect\citeauthoryear{{Ishibashi} \& {Fabian}}{{Ishibashi} \&
  {Fabian}}{2015}]{Ishibashi_Fabian_2015}
{Ishibashi} W.,  {Fabian} A.~C.,  2015, \mnras, 451, 93

\bibitem[\protect\citeauthoryear{{Ishibashi} \& {Fabian}}{{Ishibashi} \&
  {Fabian}}{2018}]{Ishibashi_Fabian_2018}
{Ishibashi} W.,  {Fabian} A.~C.,  2018, \mnras, 481, 4522

\bibitem[\protect\citeauthoryear{{Ishibashi}, {Fabian} \&
  {Maiolino}}{{Ishibashi} et~al.}{2018}]{Ishibashi_et_2018}
{Ishibashi} W.,  {Fabian} A.~C.,    {Maiolino} R.,  2018, \mnras, 476, 512

\bibitem[\protect\citeauthoryear{{Ishibashi}, {Fabian} \&
  {Reynolds}}{{Ishibashi} et~al.}{2019}]{Ishibashi_et_2019}
{Ishibashi} W.,  {Fabian} A.~C.,    {Reynolds} C.~S.,  2019, \mnras, 486, 2210

\bibitem[\protect\citeauthoryear{{King} \& {Pounds}}{{King} \&
  {Pounds}}{2015}]{King_Pounds_2015}
{King} A.,  {Pounds} K.,  2015, \araa, 53, 115

\bibitem[\protect\citeauthoryear{{King} \& {Pringle}}{{King} \&
  {Pringle}}{2006}]{King_Pringle_2006}
{King} A.~R.,  {Pringle} J.~E.,  2006, \mnras, 373, L90

\bibitem[\protect\citeauthoryear{{Kormendy} \& {Ho}}{{Kormendy} \&
  {Ho}}{2013}]{Kormendy_Ho_2013}
{Kormendy} J.,  {Ho} L.~C.,  2013, \araa, 51, 511

\bibitem[\protect\citeauthoryear{{Krumholz} \& {Thompson}}{{Krumholz} \&
  {Thompson}}{2013}]{Krumholz_Thompson_2013}
{Krumholz} M.~R.,  {Thompson} T.~A.,  2013, \mnras, 434, 2329

\bibitem[\protect\citeauthoryear{{Leftley}, {H{\"o}nig}, {Asmus}, {Tristram},
  {Gandhi}, {Kishimoto}, {Venanzi} \& {Williamson}}{{Leftley}
  et~al.}{2019}]{Leftley_et_2019}
{Leftley} J.~H.,  {H{\"o}nig} S.~F.,  {Asmus} D.,  {Tristram} K. R.~W.,
  {Gandhi} P.,  {Kishimoto} M.,  {Venanzi} M.,    {Williamson} D.~J.,  2019,
  \apj, 886, 55

\bibitem[\protect\citeauthoryear{{Li}, {Zimmerman}, {Narayan} \&
  {McClintock}}{{Li} et~al.}{2005}]{Li_et_2005}
{Li} L.-X.,  {Zimmerman} E.~R.,  {Narayan} R.,    {McClintock} J.~E.,  2005,
  \apjs, 157, 335

\bibitem[\protect\citeauthoryear{{Mattsson}}{{Mattsson}}{2020}]{Mattsson_2020}
{Mattsson} L.,  2020, \mnras, 491, 4334

\bibitem[\protect\citeauthoryear{{Menci}, {Fiore}, {Feruglio}, {Lamastra},
  {Shankar}, {Piconcelli}, {Giallongo} \& {Grazian}}{{Menci}
  et~al.}{2019}]{Menci_et_2019}
{Menci} N.,  {Fiore} F.,  {Feruglio} C.,  {Lamastra} A.,  {Shankar} F.,
  {Piconcelli} E.,  {Giallongo} E.,    {Grazian} A.,  2019, \apj, 877, 74

\bibitem[\protect\citeauthoryear{{Micha{\l}owski}}{{Micha{\l}owski}}{2015}]{Michalowski_2015}
{Micha{\l}owski} M.~J.,  2015, \aap, 577, A80

\bibitem[\protect\citeauthoryear{{Murray}, {M{\'e}nard} \& {Thompson}}{{Murray}
  et~al.}{2011}]{Murray_et_2011}
{Murray} N.,  {M{\'e}nard} B.,    {Thompson} T.~A.,  2011, \apj, 735, 66

\bibitem[\protect\citeauthoryear{{Murray}, {Quataert} \& {Thompson}}{{Murray}
  et~al.}{2005}]{Murray_et_2005}
{Murray} N.,  {Quataert} E.,    {Thompson} T.~A.,  2005, \apj, 618, 569

\bibitem[\protect\citeauthoryear{{Owen} \& {Barlow}}{{Owen} \&
  {Barlow}}{2015}]{Owen_Barlow_2015}
{Owen} P.~J.,  {Barlow} M.~J.,  2015, ArXiv e-prints

\bibitem[\protect\citeauthoryear{{Pfrommer}, {Pakmor}, {Schaal}, {Simpson} \&
  {Springel}}{{Pfrommer} et~al.}{2017}]{Pfrommer_et_2017}
{Pfrommer} C.,  {Pakmor} R.,  {Schaal} K.,  {Simpson} C.~M.,    {Springel} V.,
  2017, in 6th International Symposium on High Energy Gamma-Ray Astronomy
  Vol.~1792 of American Institute of Physics Conference Series, {Cosmic ray
  feedback in galaxies and active galactic nuclei}.
p. 030003

\bibitem[\protect\citeauthoryear{{Proga}, {Stone} \& {Kallman}}{{Proga}
  et~al.}{2000}]{Proga_et_2000}
{Proga} D.,  {Stone} J.~M.,    {Kallman} T.~R.,  2000, \apj, 543, 686

\bibitem[\protect\citeauthoryear{{Reynolds}}{{Reynolds}}{2019}]{Reynolds_2019}
{Reynolds} C.~S.,  2019, Nature Astronomy, 3, 41

\bibitem[\protect\citeauthoryear{{Roth}, {Kasen}, {Hopkins} \&
  {Quataert}}{{Roth} et~al.}{2012}]{Roth_et_2012}
{Roth} N.,  {Kasen} D.,  {Hopkins} P.~F.,    {Quataert} E.,  2012, \apj, 759,
  36

\bibitem[\protect\citeauthoryear{{Sarangi}, {Dwek} \& {Kazanas}}{{Sarangi}
  et~al.}{2019}]{Sarangi_et_2019}
{Sarangi} A.,  {Dwek} E.,    {Kazanas} D.,  2019, \apj, 885, 126

\bibitem[\protect\citeauthoryear{{Sugimura}, {Hosokawa}, {Yajima} \&
  {Omukai}}{{Sugimura} et~al.}{2017}]{Sugimura_et_2017}
{Sugimura} K.,  {Hosokawa} T.,  {Yajima} H.,    {Omukai} K.,  2017, \mnras,
  469, 62

\bibitem[\protect\citeauthoryear{{Sun} \& {Malkan}}{{Sun} \&
  {Malkan}}{1989}]{Sun_Malkan_1989}
{Sun} W.-H.,  {Malkan} M.~A.,  1989, \apj, 346, 68

\bibitem[\protect\citeauthoryear{{Takeo}, {Inayoshi}, {Ohsuga}, {Takahashi} \&
  {Mineshige}}{{Takeo} et~al.}{2018}]{Takeo_et_2018}
{Takeo} E.,  {Inayoshi} K.,  {Ohsuga} K.,  {Takahashi} H.~R.,    {Mineshige}
  S.,  2018, \mnras, 476, 673

\bibitem[\protect\citeauthoryear{{Thompson}, {Fabian}, {Quataert} \&
  {Murray}}{{Thompson} et~al.}{2015}]{Thompson_et_2015}
{Thompson} T.~A.,  {Fabian} A.~C.,  {Quataert} E.,    {Murray} N.,  2015,
  \mnras, 449, 147

\bibitem[\protect\citeauthoryear{{Vasudevan}, {Fabian}, {Reynolds}, {Aird},
  {Dauser} \& {Gallo}}{{Vasudevan} et~al.}{2016}]{Vasudevan_et_2016}
{Vasudevan} R.~V.,  {Fabian} A.~C.,  {Reynolds} C.~S.,  {Aird} J.,  {Dauser}
  T.,    {Gallo} L.~C.,  2016, \mnras, 458, 2012

\bibitem[\protect\citeauthoryear{{Wesson}, {Barlow}, {Matsuura} \&
  {Ercolano}}{{Wesson} et~al.}{2015}]{Wesson_et_2015}
{Wesson} R.,  {Barlow} M.~J.,  {Matsuura} M.,    {Ercolano} B.,  2015, \mnras,
  446, 2089

\bibitem[\protect\citeauthoryear{{Williamson}, {H{\"o}nig} \&
  {Venanzi}}{{Williamson} et~al.}{2019}]{Williamson_et_2019}
{Williamson} D.,  {H{\"o}nig} S.,    {Venanzi} M.,  2019, \apj, 876, 137

\bibitem[\protect\citeauthoryear{{Ysard}, {Jones}, {Demyk}, {Bout{\'e}raon} \&
  {Koehler}}{{Ysard} et~al.}{2018}]{Ysard_et_2018}
{Ysard} N.,  {Jones} A.~P.,  {Demyk} K.,  {Bout{\'e}raon} T.,    {Koehler} M.,
  2018, \aap, 617, A124

\bibitem[\protect\citeauthoryear{{Zhang} \& {Davis}}{{Zhang} \&
  {Davis}}{2017}]{Zhang_Davis_2017}
{Zhang} D.,  {Davis} S.~W.,  2017, \apj, 839, 54

\bibitem[\protect\citeauthoryear{{Zubovas}}{{Zubovas}}{2018}]{Zubovas_2018}
{Zubovas} K.,  2018, \mnras, 473, 3525

\bibitem[\protect\citeauthoryear{{Zubovas} \& {King}}{{Zubovas} \&
  {King}}{2012}]{Zubovas_King_2012}
{Zubovas} K.,  {King} A.,  2012, \apjl, 745, L34

\bibitem[\protect\citeauthoryear{{Zubovas} \& {King}}{{Zubovas} \&
  {King}}{2019}]{Zubovas_King_2019}
{Zubovas} K.,  {King} A.,  2019, \mnras, 489, 1373

\end{thebibliography}

\label{lastpage}

\end{document}